\documentstyle[12pt]{article}
\begin{document}
\begin{titlepage}

\begin{center}
{\large {\bf HYPERK\"AHLER METRICS BUILDING\\ AND INTEGRABLE
MODELS}}\\[0pt] \vspace{2cm} {\bf E.H. SAIDI $^{1}$ and M.B.
SEDRA$^{2}$} \\[0pt]
$^{1,\, 2}$ {\small \ International Centre for Theoretical Physics, Trieste, Italy.}\\[%
0pt]
$^{1,\, 2}${\small \ Virtual African Center For Basic Sciences and Technology, VACBT,\\[%
0pt] Focal point: Lab/UFR-Physique des Haute Energies, Facult\'{e}
des Sciences, Rabat, Morocco}.\\[0pt] $^{1,\, 2}${\small \
Groupement National de Physique des Hautes Energies, GNPHE, Rabat,
Morocco,}\\[0pt] {\small \ \ \ } {\small \ \ \ }$^{2}${\small \
Universit\'{e} Ibn Tofail, Facult\'{e} des Sciences,
D\'{e}partement de Physique,}\\[0pt] {\small \ \ \ }{\small \
Laboratoire de Physique de la Mati\`{e}re et Rayonnement (LPMR),
K\'{e}nitra, Morocco}\\[0pt]
\end{center}

\centerline{ABSTRACT}
\bigskip

Methods developed for the analysis of integrable systems are used
to study the problem of hyperK\"ahler metrics building as
formulated in $D=2$ $N=4$ supersymmetric harmonic superspace. We
show, in particular, that the constraint equation
$\beta\partial^{++2}\omega -\xi^{++2}\exp 2\beta\omega =0$ and its
Toda like generalizations are integrable. Explicit solutions
together with the conserved currents generating the symmetry
responsible of the integrability of these equations are given.
Other features are also discussed. \vskip2truecm

\end{titlepage}

\section{Introduction}

Recently much interest has been shown in the study of nonlinear
integrable models from different points of view [1, 2]. These are exactly solvable
systems exhibiting infinite dimensional local symmetries and are involved in
various areas of mathematical physics. The aim of this paper is to draw the basic
lines of a new area where the development made in integrable theories can be
used. This area deals with the construction of new explicit hyperK\"ahler metrics of
the four dimensional euclidean gravity and their generalizations. Recall that the
problem of hyperK\"ahler metrics building is an interesting question of
hyperK\"ahler geometry that can be solved in a nice way in harmonic superspace
HS [3, 4] if one knows how to solve the following nonlinear differential
equations on the sphere $S^2$:
\begin{eqnarray}
\partial^{++}q^+-\partial^{++}\left({\partial V^{4+}\over\partial (\partial^{++}
\bar q^+)}\right) +{\partial V^{4+}\over\partial\bar q^+}  & = & 0 \nonumber \\
&&\\
\partial^{++}\bar q^{++}+\partial^{++}\left({\partial V^{4+}\over\partial
(\partial^{++}\bar q^+)}\right) - {\partial V^{4+}\over\partial q^+} & = & 0 \nonumber
\end{eqnarray}
where $q^+=q^+(z,\bar z,u^\pm )$ and its conjugates $\bar q^+=q^+(z,\bar z,u^\pm )$
are complex  fields defined on $R^2\otimes S^2\approx C\!\!\!\! I\otimes S^2$,
respectively, parametrized by  the  local  analytic coordinates  $z,\bar z$  and the
harmonic variables $u^\pm\cdot\partial^{++}=u^{+i}\ \partial/\partial u^{-i}$
is  the so--called harmonic derivative. $V^{4+}=V^{4+}(q,u)$ is an interacting
potential  depending in general on $q^+,\bar q^+$, their derivatives and the
$u^\pm$'s. As described in
[3], the fields $q^+$ and $\bar q^+$ are globally defined on the sphere $S^2\approx
SU(2)/U(1)$ and may be expanded  into an infinite series preserving the total
charge, as  shown here below:
\begin{equation}
q^+(z,\bar z,u)=u^+_i\ \varphi^i(z,\bar z)+u^+_i\ u^+_j\ u^-_k\ \varphi^{(ijk)}
(z,\bar z)+\dots
\end{equation}
Note that Eqs.(1), which fix the $u$--dependence of the $q^+$'s, is in fact the pure
bosonic projection of a two dimensional $N=4$ supersymmetric HS superfield
equation of motion [4]. The remaining equations carry the spinor contributions.
They describe among other things the space time dynamics of the physical
degrees of freedom, namely the four bosons $\varphi^i(z\bar ,z); i=1,2$ and their
$D=2$ $N=4$ supersymmetric partners. For more details see Section 2.

An equivalent way of writing Eqs.(1) is to use the Howe--Stelle--Townsend (HST)
realization of the $D=2\ N=4$ hypermultiplet $(0^4, (1/2)^4)$ [5, 3]. In this
representation, that will be used in this paper, Eqs.(1) read as
\begin{equation}
\partial^{++2}\omega-\partial^{++}\left[{\partial H^{4+}\over\partial
(\partial^{++}\omega )}\right] +{\partial H^{4+}\over\partial\omega}=0\ ,
\end{equation}
 where $\omega =\omega (z,\bar z, u)$ is a real field defined on $C\!\!\!\! I\otimes
S^2$  and  whose leading terms of  its harmonic expansion read as:
\begin{equation}
\omega (z,\bar z, u)=u^+_i\ u^-_j \ f^{ij}(z,\bar z)+u^+_i\ u^+_j\ u^-_k\ u^-_\ell\
g^{ijk\ell}(z,\bar z)+\dots
\end{equation}
Similary as for Eqs.(1), the interacting potential $H^{4+}$  depends in general on
$\omega$, its  derivatives and on the harmonics. In the remarkable case where the
potentials $V^{4+}$ and  $H^{4+}$  do not depend on the derivatives of the
fields $q^+$ and $\omega$, Eqs.(1) and  (3) reduce to
\begin{eqnarray}
\partial^{++}\ q^+ +{\partial V^{4+}\over\partial q^+} & = & 0\\
\partial^{++2}\ \omega +{\partial H^{4+}\over\partial\omega} & = & 0\ .
\end{eqnarray}
Remark that the solutions of these equations depend naturally on the potentials
$V^{4+}$   and  $H^{4+}$  and then the finding of these solutions is not an easy
question.  There are only few examples that had been solved exactly. The first
example we  give is the Taub--Nut model leading to the well--known Taub--Nut metric of
the  four dimensional euclidean gravity. Its potential $V^{4+}(q^+,\bar q^+)$ is given
by [4]
\begin{equation}
V^{4+}={\lambda\over 2}\ (q^+\bar q^+)^2\ ,
\end{equation}
where $\lambda$ is a real coupling constant. Putting Eq.(7) back into Eq.(5), one gets
\begin{equation}
\partial^{++}\ q^+ +\lambda (q^+\bar q^+)\bar q^+=0
\end{equation}
whose solution reads as [4]
\begin{equation}
q^+(z,\bar z,u)=u^+_i\ \varphi^i(z,\bar z)\exp -\lambda (u^+_ku^-_\ell
\varphi^k\bar\varphi^\ell )\ .
\end{equation}
Following Ref.[4], the knowledge of the solution Eq.(9) of Eq.(8) is the key point in
the  identification of the metric of the manifold parametrized by the bosonic fields
$\varphi^i(z,\bar z)$ and $\bar\varphi^i(z,\bar z)$ of the $D=2\ N=4$
supersymmetric nonlinear Taub--Nut
$\sigma$--model whose bosonic part reads as [4]
$$
S^{TN}_B=-{1\over 2}\int dz\ d\bar z\left( g_{ij} \partial_z \varphi^i
\partial_{\bar z} \varphi^j+\bar g^{ij} \partial_z \bar\varphi_i \partial_{\bar
z} \bar\varphi_j+2h^i_j \partial_z \varphi^j \partial_{\bar
z}\bar\varphi_i\right)
$$
where
\begin{eqnarray}
g_{ij} & = & {\lambda (2+\lambda f\bar f)\over 2(1+\lambda f\bar f)}\ \bar f_i\bar
f_j,\quad \bar g^{ij}={\lambda (2+f\bar f)\over 2(1+\lambda f\bar f)}\ f^if^j
\nonumber \\
h^i_j & = & \delta^i_j(1+\lambda f\bar f) -{\lambda (2+\lambda f\bar f)\over
2(1+\lambda f\bar f)}\ f^i\bar f_j,\\
f\bar f & = & f^i\bar f_i \ .\nonumber
\end{eqnarray}
Moreover, using the HST representation of the $D=2\ N=4$ hypermultiplet, Eq.(8)
may be rewritten as
\begin{equation}
\left[ \partial^{++}+\lambda\
{\omega\mathop{\partial}\limits^{\leftrightarrow}{^{++}}\ \bar\omega\over
(1+2\omega\bar\omega )}\right]^2\cdot\omega =0\ ,
\end{equation}
where $\bar\omega$ is the complex conjugate of $\omega$ and
$\omega\mathop{\partial}\limits^{\leftrightarrow}{^{++}}\
 \bar\omega =\omega\partial\bar\omega -\partial\omega\cdot\bar\omega$. Here also,
this equation is exactely solvable. The solution reads as
\begin{equation}
\omega (z,\bar z, u)=u^+_i\ u^-_j\ f^{ij}(z,\bar z)\exp -i\lambda\beta
\end{equation}
where
\begin{equation}
\beta =u^+_k\ u^-_\ell \left[ f^0\bar f^{(k\ell )}-\bar f^0f^{(ij)}+\in_{rs}\
f^{(ks)}\ \bar f^{(\ell r)}\right]\ .
\end{equation}
The explicit form of the Taub--Nut metric in the $\omega$--representation is worked
out in Ref.[6].

The second example that has been solved exactly is the Eguchi--Hanson model
whose potential reads as [7]
\begin{equation}
H^{4+}(\omega ) =\left[ u^+_i\ u^+_j\ \xi^{(ij)}\right]^2/\omega^2\ ,
\end{equation}
where $\xi^{ij}$ is an $SU(2)$ real constant triplet. Details are exposed in [7]. It
is  interesting to note here  that for the above mentioned nonlinear differential
equations and their generalizations, the integrability is due to the existence of
symmetries allowing their linearizations.

In this paper, we focus our attention on Eq.(6) and look for potentials leading to
exact solutions of this equation. Our method is based on suggesting new plausible
integrable equations by proceeding by formal analogy with the known integrable two
dimensional nonlinear differential equations especially the Liouville equation and
its  Toda generalizations.

Among our results we show that the model described by the potential
\begin{equation}
H^{4+}(\omega ,u)=-{1\over 2}\ \left({\xi^{++}\over\lambda}\right)^2\exp
2\lambda\omega
\end{equation}
which implies in turn the following equation
\begin{equation}
\lambda\ \partial^{++2}\omega -\xi^{++2}\exp 2\lambda\omega =0
\end{equation}
is integrable. The explicit solution of this nonlinear differential equation reads as
\begin{equation}
\xi^{++}\exp\lambda\omega ={u^+_i\ u^+_j\ f^{ij}(z,\bar z)\over 1-u^+_k\ u^-_\ell\
f^{k\ell}(z,\bar z)}\ .
\end{equation}
Here also we show that the integrability of Eq.(16) is due to the existence of a
symmetry generated by the following conserved current
\begin{eqnarray}
t^{4+} & = & (\partial^{++}\omega )^2-{1\over\lambda}\ \partial^{++2}\omega
\nonumber \\
\\
\partial^{++}\ t^{4+} & = & 0\ . \nonumber
\end{eqnarray}
This representation of the current $t^{4+}$, which in some sense resembles to the
Liouville current, can also be obtained by using the field theoretical method or
again with the help of an extended Miura transformation which reads, in the general
situation, as
\begin{equation}
(\partial^{++n}-W^{2n+})=\mathop{\Pi}\limits^n_{j=1}\ (\partial^{++} -V^{++}_j)\ ,
\end{equation}
where the fields $V^{++}_j, j = 1,\dots, n$, obeys the traceless condition namely
$\mathop{\sum}\limits^n_{j=1}\ V^{++}_j=0$. The extension of Eqs.(15)--(16) and (18)
for a rank $n$ simple Lie algebra is also studied.

The presentation of this paper is as follows: First we review briefly the
hyperK\"ahler metrics building from the HS method and show that the solving of
the pure bosonic Eq.(6) is the main step to achieve in this programme. Then,
we present our integrable model and discuss its Toda like generalizations. After
that we introduce the generalized Miura transformation and derive a series of
conserved currents $W^{2n+}, n\geq 1$ responsible for the integrability of the
generalizations of our model. Finally,  we give our conclusion.

\section{Generalities on the hyperK\"ahler metrics building from HS}

HyperK\"ahler metrics have vanishing Ricci tensors and are then natural
solutions of the Einstein equation in the vacuum [8]. They also appear as the
metrics of the bosonic manifolds of two dimensional $N=4$ supersymmetric nonlinear
$\sigma$--models describing the self coupling of the hypermultiplet $(0^4, (1/2)^4)$
[9, 4, 5, 7]. A powerful method to write down the general form of such models, and
then build the underlying hyperK\"ahler metrics, is given by harmonic superspace.  The
latter is parametrized by the supercoordinates $Z^M
=(z^M_A,\theta^-_r,\bar\theta^-_r)$ where $z^M_A =  (z,\bar
z,\theta^+_r,\bar\theta^+_r,u^\pm )$ are the supercoordinates of the so--called
analytic subspace in  which $D=2\ N=4$ supersymmetric theories are formulated [3].
$d^2zd^4q\theta^+du$ is the HS  integral measure in the $z^M_A$  basis. The matter
superfield $(0^4, (1/2)^4)$ is realized by  two dual analytic superfields $Q^+ =
Q^+(z,\bar z,\theta^+, \bar\theta^+, u)$ and $\Omega =\Omega (z,\bar z,
\theta^+,\bar\theta^+, u)$  whose leading bosonic fields are respectively given by
$q^+$ and $\omega$ Eqs.(2) and (4).

The action describing the general coupling of the analytic superfield $\Omega$ we
are interested in here, reads as [3]:
\begin{equation}
S[\Omega ]=\int d^2z\ d^4\theta^+\ du\left({1\over 2}\ (D^{++}\Omega
)^2-H^{4+}(\Omega ,u)\right)\ ,
\end{equation}
where the harmonic derivative $D^{++}$  is given by
\begin{equation}
D^{++}=\partial^{++}-2\bar\theta^+_r\ \theta^+_r\ \partial_{-2r}\ .
\end{equation}
The superfield equation of motion
\begin{equation}
D^{++2}\Omega -D^{++}\left({\partial H^{4+}\over \partial D^{++}\Omega}\right) +
{\partial H^{4+}\over\partial\Omega}=0
\end{equation}
which depends naturally on the interaction, contains as many differential
equations as the component fields carried by the superfield $\Omega$. It turns out
that  this system of differential equations splits into three subsets of equations
depending on  the canonical dimensions of the component fields of the superfield
$\Omega$. To be  more precise let us describe briefly the main steps of the
hyperK\"ahler metrics   building  procedure.\\
1. Specify the self interacting potential $H^{4+}(\Omega , u)$, since to each
interaction  corresponds a definite metric. As an example, one may take the
Taub--Nut  interaction $\lambda (Q^+\bar Q^+)^2$ [4], or again the Eguchi--Hanson
one: $(\xi^{++}/\Omega )^2$ [7]. These  potentials do not depend on the superfield
derivatives and lead to equations type  (5) and (6). Another interesting example
that will be considered in this study is  given by:
\begin{equation}
H^{4+}(\Omega ,u^+)=-{1\over 2}\ (\xi^{++}/\beta)^2\exp 2\beta\Omega\ ,
\end{equation}
where $\beta$ is a coupling constant and $\xi^{++}=u^+_i\ u^+_j\ \xi^{(ij)}$, a
constant isotriplet similar to that appearing in the Eguchi--Hanson model.\\
2. Write down the corresponding superfield equation of motion, namely
\begin{equation}
D^{++2}\ \Omega +{\partial H^{4+}\over\partial\Omega}=0\ ,
\end{equation}
which reads, for the potential (23), as:
\begin{equation}
\beta\ D^{++2}\ \Omega-(\xi^{++})^2\exp 2\beta\Omega =0\ .
\end{equation}
3. Expanding the analytic superfield $\Omega$ in $\theta^+_r$ and $\theta^+_r$ series
as
\begin{eqnarray}
\Omega & = & \omega +(\theta^+_-\ \theta^+_+\
F^{--}+\bar\theta^+_+\ \bar\theta^+_-\ \bar
F^{--})+(\bar\theta^+_-\ \theta^+_+\ G^{--}+\nonumber \\
&&\bar\theta^+_+\ \theta^+_-\ \bar
G^{--}+(\bar\theta^+_-\
\theta^+_-\ B^{--}_{++}+\bar\theta^+_+\ \theta^+_+\ B^{--}_{--}+\nonumber \\
&&\bar\theta^+_-\ \theta^+_-\ \bar\theta^+_+\ \theta^+_+\ \Delta^{(-4)}\ ,
\end{eqnarray}
where we have set the spinor fields to zero for simplicity. Then
putting back into  Eq.(24) one obtains three kinds of differential equations: The
first one which is  given by the equation of motion of the Lagrange field
$\Delta^{(-4)}$ of canonical  dimension 2, reads as:
\begin{equation}
\partial^{++2}\ \omega+{\partial H^{4+}(\omega)\over\partial\omega} =0\ .
\end{equation}
This is a constraint equation fixing the dependence of $\omega$ in terms of the free
bosonic fields $f^{ij}$ of the $D=2\ N=4$ hypermultiplet. The knowledge of the
solution  of this nonlinear equation is necessary as it is one of the two main
difficult steps  in the construction of hyperK\"ahler metrics in this way. In the next
section, we  shall show that the methods developed in integrable theories can be used
to  solve a specific class of these equations such that the equation implied by the
potential Eq.(23)
\begin{equation}
\beta\ \partial^{++2}\omega-\xi^{++2}\exp 2\beta\omega =0
\end{equation}
or again its generalizations. The second set of relations are given by the equations
of motion of the auxiliary fields $F^{--}, G^{--}$, and $B^{--}_{rr}$ of canonical
dimensions one. In  all known cases, the solutions of these equations are obtained by
making  an appropriate change of variables inspired from the solution of Eq.(27). Note
that for  the potential (25), the auxiliary fields equations read as
\begin{eqnarray}
\beta\ \partial^{++2}\ F^{--} & - &\xi^{++2}\ F^{--}\exp 2\beta\omega  =  0 \nonumber
\\
\beta\ \partial^{++2}\ G^{--} & - &\xi^{++2}\ G^{--}\exp 2\beta\omega = 0 \\
\beta\ \partial^{++2}\ B^{--}_{rr} & - &\xi^{++2}\ B^{--}_{rr}\exp 2\beta\omega  =
4\partial^{++}\ \partial_{rr}\omega\ . \nonumber
\end{eqnarray}
These equations, when solved, give the relation between the auxiliary fields
$F^{--}, G^{--}$, and  $B^{--}_{rr}$  and the $D=2$ matter field $\omega$ and its
space--time derivatives.  Note also that Eqs.(27)--(29) once integrated fix $D=2\
N=4$ supersymmetry partially  on shell. The last equation is given by the space--time
equation of motion of $\omega$. It  describes the dynamics of $\omega$ and is not
involved in the hyperKahler metric  building programme.\\
4. Combining the results of the steps one, two and three, one finds that the
bosonic part of the action Eq.(20) takes a form similar to Eq.(10) from which one
can read the hyperK\"ahler metric directly.\\
At the end of this section we would like to point out that, in general,  the energy
momentum tensor of the action (20) reads in terms of the real superfield $\Omega$ and
the interacting potential as:
\begin{equation}
T^{4+}(\Omega ) ={1\over 2}(D^{++}\Omega )^2-{\partial H^{4+}\over
\partial D^{++}\Omega}\cdot D^{++}\Omega -H^{4+}\ .
\end{equation}
The conservation law of this current can easily be checked with the help of the
equation  of motion (22). For the interacting potential given by Eq.(23), the above
conserved  current takes the remarkable form
\begin{equation}
T^{4+}(\Omega )={1\over 2}(D^{++}\Omega )^2-{1\over\beta}\ D^{++2}\Omega\ .
\end{equation}

\section{The integrability of  $\beta\partial^{++2}\omega -\xi^{++2}\exp
2\beta\omega =0$}

In this section we prove that this nonlinear harmonic differential
equation is solvable though apparently it shows no special symmetry. The
property of solvability of this equation is expected from its formal analogy with
the well--known Liouville equation. This why we shall start by describing the
local equation of motion of the two dimensional Liouville field $\varphi (z,\bar z)$,
namely
\begin{equation}
\beta\ \partial_z\partial_{\bar z}\ \varphi -\exp (2\beta\varphi ) = 0
\end{equation}
where $\beta$ is a real coupling constant. To study the integrability of this
equation, different techniques including the Lax method were developed. Here, we
content ourselves to recall that the explicit solution of this nonlinear equation
can be written as [10]:
\begin{equation}
\exp 2\beta\varphi = cte\cdot {F'(z)\cdot\bar F'(\bar z)\over
(1-F(z)\cdot\bar F(\bar z))^2}\ ,
\end{equation}
where $F(z)$ and $\bar F(\bar z)$ are arbitrary analytic and anti--analytic
functions,  $F'(z) =\partial F(z)$ and $\bar F'(z) = \bar\partial\bar F(\bar z)$.  As
it is well known, the integrability of the nonlinear differential Liouville equation
is due to its conformal symmetry generated  by the following classical energy
momentum tensor \begin{equation}
T_L(\varphi ) =(\partial\varphi )^2-{1\over\beta}(\partial^2\varphi )\ .
\end{equation}
The conservation law of this current follows immediately by using the equation
of motion as shown here below.
\begin{eqnarray}
\bar\partial\ T_L(\varphi ) & = & 2\Box\varphi\partial\varphi -{1\over\beta}\
\partial(\Box\varphi ) \nonumber \\
& = & -{1\over\beta}(\partial -2\beta\partial\varphi )\ \Box\varphi = 0\ .
\end{eqnarray}
Having given the necessary ingredient of the classical Liouville equation, we pass
now to  study our equation.
\begin{equation}
\beta\ \partial^{++2}\omega -\xi^{++2}\exp 2\beta\omega =0\ .
\end{equation}
This is a nonlinear harmonic differential equation which in principle, is not easy
to solve. However, forgetting about the constant $\xi^{++}$ and the global properties
of  the field $\omega$ with respect to the coordinates of $S^2$, Eq.(16) shows a
striking  resemblence with the Liouville equation examined earlier. Therefore, one
should  expect that both these equations would share some general features and more
particularly  their integrabilities. Using this formal analogy with the Liouville
equation, it is not  difficult to see that the solution of Eq.(16) can be expressed in
terms of the four  physical real bosonic fields $f^{ij}$ of the $D=2\ N=4$ HST free
hypermultiplet $(0^4, (1/2)^4)$  as follows:
\begin{equation}
\xi^{++}\exp\beta\omega ={u^+_i\ u^+_j\ f^{ij}\over 1-u^+_i\ u^-_j\ f^{ij}}\ ,
\end{equation}
To check that Eq.(37) is indeed the solution of Eq.(16), note first of all that
we  have
\begin{equation}
\beta\ \partial^{++}\omega ={u^+_i\ u^+_j\ f^{ij}\over 1-u^+_k\ u^-_\ell\
f^{k\ell}}\ ,
\end{equation}
which with the help of Eq.(37), it also reads as:
\begin{equation}
\beta\ \partial^{++}\omega =\xi^{++}\exp\beta\omega\ .
\end{equation}
Acting on this relation by the harmonic differential operator $\partial^{++}$, we get
after  setting $f^{++}=u^+_iu^+_jf^{ij}$ and $f=u^+_iu^-_jf^{ij}$ for simplicity:
\begin{equation}
\beta\ \partial^{++2}\omega =\left({f^{++}\over 1-f}\right)^2\ .
\end{equation}
Using Eq.(37) once again, one sees that Eq.(40) can be rewritten as
\begin{equation}
\beta\ \partial^{++2}\omega =\xi^{++2}\exp 2\beta\omega\ ,
\end{equation}
which coincides exactly with Eq.(16). Note by the way that the solution
(37) looks, in some  sense, like that given by Eq.(33). Thus the question is, can we
find appropriate symmetry responsible of the integrability of Eq.(16)? The answer to
this question is obtained by exploiting once  more the analogy with the symmetry of
the Liouville equation (32). There, the  symmetry is the conformal invariance
generated by the conserved current Eq.(34).  In our case the classical current
generating the symmetry of Eq.(16) is expected to  have the following natural form:
\begin{equation} t^{4+}(\omega )=(\partial^{++}\omega )^2-1/\beta\
\partial^{++2}\omega \end{equation}
which is just the pure bosonic projection of Eq.(31). Here also the conservation
law of this current follows by using Eq.(16). Indeed we have
\begin{eqnarray}
\partial^{++}\ t^{4+} & = & 2(\partial^{++}\omega )(\partial^{++2}\omega )-1/\beta\
\partial^{++3}\omega \nonumber \\
& &\\
& = & -1/\beta^2 (\partial^{++}-2\beta\ \partial^{++}\omega )\beta\
\partial^{++2}\omega\ , \nonumber
\end{eqnarray}
which vanishes identically with the help of the identity (41). \\
Furthermore, knowing that integrable two dimensional field theoretical models are
intimately related to the  simple roots system of Lie algebras, our goal in the
next discussion is to use this crucial property to test our expected integrable
model Eq.(40).  The Liouville equation discussed at the beginning of this section is
in fact the leading  case of a system of integrable equations describing the
so--called  $A_n$--Toda models with  $n= 1, 2,\dots$. Denoting by
$\{\vec\alpha_i;1\leq i\leq n\}$ the simple roots system of the $A_n$--Lie algebra
and by \begin{equation} \vec\omega =\sum^n_{i=1}\ \vec\alpha_i\ \omega_i\ ,
\end{equation} an $O(n)$ vector of $n$ component fields $\omega_i$, the $A_n$--Toda
like extension of Eq.(36)  reads as:
\begin{equation}
\beta\ \partial^{++2}\vec\omega -\xi^{++2}\ \sum^n_{j=1}\
\vec\alpha_i\exp\beta\vec\alpha_j\cdot\vec\omega =0\ .
\end{equation}
The integrability of these equations is ensured by showing the existence of $n$
independent conserved currents type Eq.(42) generating their underlying
symmetries. This will be done in the next section. We end this study by noting
that Eq.(45) is just the pure bosonic projection of a superfield equation of motion
\begin{equation}
\beta\ D^{++2}\Omega -\xi^{++2}\ \sum^n_{j=1}\ \vec\alpha_j\exp\beta\vec\alpha_j\
\vec\Omega\ ,
\end{equation}
obtained by variating the following HS superspace action
\begin{equation}
S[\Omega ]=\int d^2z\ d^2\theta^+\left\{{1\over 2}\ D^{++}\vec\Omega\cdot
D\vec\Omega -(\xi^{++}/\beta )^2\
\sum^n_{i=1}\exp\beta\vec\alpha_i\vec\Omega\right\}\ .
\end{equation}

\section{Generalized Miura transformation}

We start by recalling that for bosonic Toda conformal field theories  based
on the simple Lie algebra $A_n$, the field realization of the higher spin currents is
given by the so--called $WA_n$ Miura transformation  namely [11]:
\begin{equation}
\partial^n_z-\sum^n_{k=2}\ u_k\ \partial^{n-k}=\mathop{\Pi}\limits^n_{j=1}\
(\partial_z-q_z)_j\ ,
\end{equation}
where  the $q_{zj}$'s, $j = 1,\dots, n$, are spin one analytic fields
obeying $\mathop{\sum}\limits^n_{j=1}\ q_{zi}=0$ and  where we have used the
convention notation:
\begin{equation}
(\partial_z-q_z)_j=(\partial_z-q_{zj})\ ,
\end{equation}
Expanding the r.h.s of Eq.(48), one gets the field realization of the conformal spin
$k$  conserved currents  $u_k$ . Naturally this method applies for super TCFT's as
well  [11]. In our present case, one can define an adapted Miura transformation by
generalizing the analysis of the beginning of Section 3 based on the formal analogy
between the Liouville theory and Eq.(16). This transformation, to which we shall
refer hereafter to as the generalized Miura transformation, reads in the language
of HS  superfields as:
\begin{equation}
(D^{++})^n-\sum^n_{k=2}\ J^{+2k}(D^{++})^{n-k}=\mathop{\Pi}\limits^n_{j=1}\
(D^{++}-V^{++})_j\ ,
\end{equation}
where
\begin{eqnarray}
(D^{++}-V^{++})_j & = & D^{++}-V^{++}_j \nonumber \\
&&\\
\sum^n_{j=1}\ V^{++}_j & = &0\ . \nonumber
\end{eqnarray}
Expanding the r.h.s of Eq.(50), one naturally gets the superfield realization of the
conserved current $J^{+2k}$. Let us describe briefly hereafter the $n=2$ and 3
situations. In the first case, Eq.(50) leads to the following realization of the
unique current $J^{4+}$
\begin{equation}
J^{4+}=(V^{++})^2-D^{++}\ V^{++}
\end{equation}
where we have used $V^{++}=V^{++}_2=-V^{++}_1$. Taking $V^{++}=\beta
D^{++}\Omega$ and
$J^{++}=\beta^2T^{4+}$, the above equation reduces to Eq.(31) giving the energy
momentum  tensor whose pure bosonic projection coincides with Eq.(42). For
the  $n=3$ case, the generalized Miura transformation leads to the following
superfield  realizations of the two supercurrents $J^{4+}$  and   $J^{6+}$:
\begin{eqnarray}
J^{4+} & = & \left[ (V^{++}_1)^2+2D^{++}V^{++}_1\right] + \left[
(V^{++}_2)^2+D^{++}V^{++}_2\right] +V^{++}_1\cdot V^{++}_2 \nonumber \\
J^{6+} & = & -(V^{++}_1V^{++}_2)(V^{++}_1+V^{++}_2)+(V^{++}_1+V^{++}_2)
D^{++}V^{++}_1 \nonumber \\
&& -D^{++}(V^{++}_1V^{++}_2)+D^{++2}V^{++}_1
\end{eqnarray}
where $V^{++}_1$ and $V^{++}_2$ are two free superfields. Setting $V^{++}_i=\beta\
D^{++}\Omega_i$, the  conservation laws of these currents follows as usual with the
help of the equations of  motion (46).

\section{Conclusion}

In this paper, we have shown that the two dimensional integrable model
techniques may be used in the accomplishment of the programme of
constructing of new hyperK\"ahler metrics. Recall once again that the problem of
hyperK\"ahler metrics building can be studied in a convenient way in harmonic
superspace.  The main difficulty in this approach is the solving of nonlinear
harmonic  differential equations. Only few and special examples such as Taub--Nut
and   Eguchi--Hanson models and some of their generalizations were solved exactly in
 literature. In all these cases, the key point in solving these equations is the
existence of a symmetry allowing their linearization. In the present study, using a
formal analogy with TCFT's, we have succeeded to draw the main lines of a new
class of nonlinear two dimensional $N=4$ supersymmetric $\sigma$--models that can be
solved exactly. These models are described by the following HS action
\begin{equation}
S[\Omega ]=\int d^2z\ d^4\theta^+du\left({1\over 2}\ D^{++}\vec\Omega\cdot
D^{++}\vec\Omega -(\xi^{++}/\beta )^2\sum^n_{i=1}\exp\vec\alpha_i\vec\Omega\right)
\end{equation}
where $\vec\alpha_i, i = 1,\dots, n$,  are the simple roots of a rank $n$ simple Lie
algebra.
$\vec\Omega =\mathop{\sum}\limits^n_{i=1}\ \vec\alpha_i\Omega_i$  and
$\xi^{++}=u^+_ku^+_\ell\xi^{(k\ell )}, k\ell = 1,2,$  is a constant. In the general
case we  have shown that the pure bosonic projection of the superfield equation of
motion,  namely
\begin{equation}
\beta\ \partial^{++2}\vec\omega -(\xi^{++})^2\ \sum^n_{i=1}\ \vec\alpha_i\exp\beta\
\vec\alpha_i\vec\omega =0
\end{equation}
is integrable. The corresponding $(n-1)$ conserved current, responsible of the
integrability of the above equation, was obtained by using an adapted Miura
transformation. Here it is interesting to note that a Lax formalism similar to that
used in TCFT's can be defined also in our case. For the special case $n=2$, we have
 also given the explicit solution of Eq.(16). What remains to do is to write down
the explicit form of the underlying hyperK\"ahler metric associated to the potential
Eq.(23). This technical problem will be addressed in a future occasion.

\vskip1.5truecm

\noindent{\bf Acknowledgments}
\bigskip

One of the authors (M.B.S.) would
like to thank Professor Abdus Salam, the International Atomic Energy Agency and
UNESCO for hospitality at the International Centre for Theoretical Physics, Trieste.
He would also like to thank Professor S. Randjbar-Daemi for his hospitality and
scientific help.

\newpage

\centerline{REFERENCES}
\begin{description}

\item{[1]}
See for instance:
    B. Kupershmidt, Integrable and Superintegrable Systems World Scientific
    (1990);\\
 A. Das, Integrable models. World Scientific (1989)

\item{[2]}
Mussardo, Phys. Rep. {\bf C218} (1992)

\item{[3]}
A. Galperin, E. Ivanov, S. Kalitzin, V. Ogievetsky and E. Sokatchev, Class.
    Quant. Grav. {\bf 1} (1984) 469

\item{[4]}
A. Galperin, E. Ivanov, V. Ogievetsky and E. Sokatchev, Com. Math. Phys.
{\bf    103} (1986) 511;\\
D. Olivier and G. Valent,  Phys. Lett. {\bf B189} (1987) 79

\item{[5]}
P. Howe, K. Stelle and P. K Townsend, Nucl. Phys. {\bf B214} (1983) 519

\item{[6]}
T. Lhallabi and E.H. Saidi,  Int. J. Mod. {\bf A4} (1989) 351

\item{[7]}
A. Galperin, E. Ivanov, V. Ogievetsky and P. K Townsend, Class. Quant.
    Grav. {\bf 3} (1986) 625

\item{[8]}
See for instance:
    A. Morozov and A. Perelemov, Phys. Rep. {\bf C146} (1987) 135 and references therein.

\item{[9]}
L. Alvarez--Gaum\'e and D. Freedman,  Com. Math. Phys. {\bf 80} (1981) 443;\\
J. Bagger and E. Witten, Nucl. Phys. {\bf B222} (1983) 1

\item{[10]}
L. Alvarez--Gaum\'e, CERN Preprint.  T.H/6123/91

\item{[11]}
J. Evans and T. Hollowood,  Nucl. Phys. {\bf B352} (1991) 723;\\
    H. Nohara,  Ann. of Phys. {\bf 214} (1992) 1
\end{description}

\end{document}